\documentclass[11pt]{article}
\usepackage{moriond,epsfig}

\def\be{\begin{equation}}
\def\ee{\end{equation}}
\def\bea{\begin{eqnarray}}
\def\eea{\end{eqnarray}}

\begin{document}
\vspace*{4cm}

\title{INTEGRABILITY OF TWIST - 3 EVOLUTION EQUATIONS IN QCD
}
\author{{A.~Manashov}}
\address{Department of Theoretical Physics, Sankt Petersburg University,
\\ 1  Ulyanovskaya street, Petergof, Sankt Petersburg, 198 904, RUSSIA}

\maketitle\abstracts{I review the recent progress in solution of the
evolution equations of the three particle hadron distribution
amplitudes.}



The three particle operators appear in QCD in the studies of the twist~--~3
parton
distributions~\cite{ABH,BBKT,BrBal}
 and   twist~--~3 meson wave functions~\cite{BF,PBBKT},
and more interestingly,
leading twist  nucleon wave functions~\cite{BLreport,CZreport,B+}.
For example, the $Q^2$ dependence of a baryon wave function is given to one loop
accuracy by the following expression
\be
\phi(x_i,Q^2)=x_1 x_2 x_3 \sum_{N,q} \phi_{N,q}\, P_{N,q}(x_i)
\left (\frac{
\alpha_s(Q)}{\alpha_s(Q_0)}
\right )^{\gamma_{N,q}/b_0},
\label{expansion}
\ee
where  $b_0=11/3\, N_c-2/3\, n_f$. 
The anomalous dimensions of the three quark operators,
$\gamma_{N,q}$,
and polynomials,  $P_{N,q}(x_i)$ arise from the solution of the eigenvalue problem for 
a certain integral operator (Hamiltonian) $H$, which effectively describes 
the dynamics of the three particle system with a pairwise interaction.
It has been observed recently~\cite{BDM} that 
for the
particular twist-3 operators
the corresponding three particle systems 
are intrisically  related to the Heisenberg spin magnet, which is known to be 
integrable quantum mechanical model and their solution can be found by
applying powerful method of Integrable Models.
\vskip 0.1cm

It is well known that the conformal symmetry of the QCD Lagrangian reveals itself
as the $SL(2,R)$ invariance of the evolutions equations~\cite{Makeenko,O82}.
As a consequence the pairwise Hamiltonians depend on the corresponding two~-~particle
Casimir operators only,
\be
H=\sum_{ik}H_{ik},\ \ \ 
H_{ik}=h(J_{ik}), \ \ \ J_{ik}(J_{ik}+1)=L^2_{ik}=(\vec{L}_i+\vec{L}_k)^2,
\ee
where $L_i$ is the $SL(2,R)$ generators associated with $i$-th particle.
Then the explicit calculations~\cite{ren3q} yield that up to 
unessential constant the Hamiltonian governing the $Q^2$ evolution
of the wave function of baryon with the maximal helicity $\lambda=3/2$
\begin{equation}
H_{3/2}=2\left(1+\frac{1}{N_c}\right)\sum_{i<k}[\psi(J_{ik}+1)-\psi(2)]
+\frac32 C_F
\end{equation}
coincides with the Hamiltonian of the noncompact $XXX_{s=-1}$ Heisenberg spin magnet.
This Hamiltonian possesses an additional integral of motion (conserved charge)
\begin{equation}
  Q = \frac{i}{2}[L_{12}^2,L_{23}^2]
= i^3 \partial_1\partial_2\partial_3 z_{12} z_{23} z_{31}\,,
\ \
\label{Q3}
\end{equation}
that commutes with $H_{3/2}$ and with $SL(2)$ generators.
It is interesting to note that this model
has already been
encountered in QCD in the studies of the Regge asymptotics of the
scattering amplitudes \cite{FK,Lip}.
The eigenvalue problem for the operator $Q$ appears to be much more simple, 
though equivalent,
those for the $H_{3/2}$ Hamiltonian, and a number of  results 
had been  obtained in
the
Refs.~\cite{BDM,BDKM,K95,K96,K97}
for their spectra. 
In particular, using the different
metods it became possible to get very accurate 
 decription of the spectrum of the
operator $Q$, and the Hamiltonian $H_{3/2}$. 
Moreover the obtained results may serve  a starting point to analysis of
the more interesting from the physical point of view case of the $\lambda=1/2$
nucleon distribution amplitudes.
\vskip 0.2cm  

\begin{figure}[t]
\centerline{\epsfxsize8cm\epsfysize6cm\epsffile{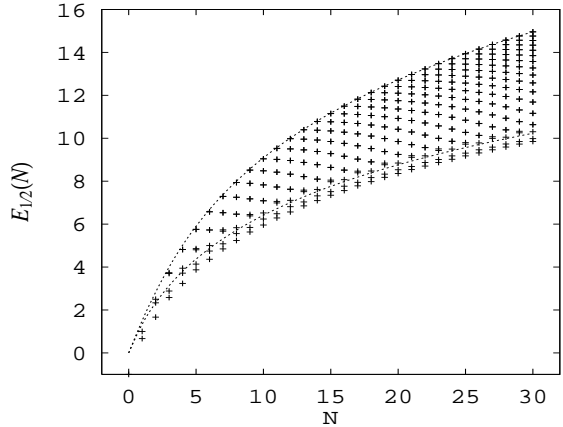}}
\caption[]{\small The spectrum of eigenvalues for the
Hamiltonian ${\cal H}_{12}$. The lines of the largest and the smallest
eigenvalues of ${\cal H}_{32}$ are indicated by  dots
for comparison.}
\label{fig:e12}
\end{figure}

The scale dependence of the $\lambda=1/2$ nucleon distribution amplitudes
is driven by the Hamiltonian
\be
H_{1/2}=H_{3/2}+V, \ \ \ \ V=-\left(L_{12}^{-2}+L_{23}^{-2}\right).
\ee
The Hamiltonian $H_{1/2}$ is not integrable and the corresponding eigenproblem
cannot be solved exactly. 
Hovewer, the operator $V$ can be treated as a small perturbation
to the Hamiltonian $H_{3/2}$ for 
the most part of the spectrum, except a few low lying levels. 
The numerical calculation  gives the spectrum shown
on Fig.~\ref{fig:e12}. The  spectra of $H_{1/2}$ and $H_{3/2}$ are
very similar in the upper part and at the same time two lowest levels of the $H_{1/2}$
Hamiltonian appears to be special and `dive' considerably below the line of the
lowest eigenvalue of $H_{3/2}$. 
The careful analysis of the low energy part of the spectrum~\footnote{
It turns out that the  resulting problem for the effective Hamiltonian
describing dynamics of the low lying levels
turns out to be a generalization of the famous Kroning-Penney model of a single
particle in a periodic delta potential~\cite{QM}.}
 allows to 
to explain this phenomenon as due to binding of the quarks with opposite helicity and
forming the scalar diquark~\cite{BDKM}.
\vskip 0.2cm

The above analysis can be extended to the case of the
twist~-~3
quark gluon operators, which had been paid much attention
recently~\cite{ABH,BBKT,BrBal,Bel1}.
The $Q^2$ evolution of the twist-3 quark gluon distribution amplitudes are driven by
the Hamiltonians (in the large $N_c$ limit)
\begin{eqnarray}
H_S&=&\psi(J_{12}+\frac52)+\psi(J_{12}-\frac12)+\psi(J_{23}+\frac32)+\psi(J_{23}+\frac12)
-\delta, \\
H_T&=&\psi(J_{12}+\frac52)+\psi(J_{12}-\frac12)+\psi(J_{23}+\frac52)+\psi(J_{23}-\frac12)
-\delta,
\end{eqnarray}
for the chiral even and chiral odd distributions, respectively, and
$\delta=4\psi(2)+3/2$.
The both Hamiltonians possess the integrals of motion~\cite{BDM} which are
\begin{eqnarray}
Q_S&=&\left\{L_{12}^2,L_{23}^2\right\}-\frac12 L_{23}^2-\frac92 L_{12}^2,\\
Q_T&=&\left\{L_{12}^2,L_{23}^2\right\}-\frac92 L_{23}^2-\frac92 L_{12}^2.
\end{eqnarray}
The underlying models can be identified as the open nonhomogeneous noncompact
spin chains. The theory of the open spin chains is less developed
as  compared with those for the closed spin chains, and many problems has
to be solved yet~(see~Ref.~\cite{DKM}).
However many results (such as calculation of the asymptotic expansion for
conserved charges and energies)
are available by the elementary methods~\cite{BDM,Bel1},
the full account can be found in Ref.~\cite{Bel1}.

\section*{Acknowledgments}
This work was carried out with the support of the
Russian Fond of Basic Research, Grant 97--01--01152.

\end{document}